\documentclass[12pt]{iopart}
\usepackage{graphicx}
\usepackage{ulem}

\begin{document}


\title{Effect of \textit{3d}-doping on the electronic structure of BaFe$_2$As$_2$}

\author{J. A. McLeod$^1$, A. Buling$^2$, R. J. Green$^1$, T. D. Boyko$^1$, N. A. Skorikov$^3$, E. Z. Kurmaev$^3$, M. Neumann$^2$, L. D. Finkelstein$^3$, N. Ni$^{4*}$, A. Thaler$^4$, S. L. Bud'ko$^4$, P. C. Canfield$^4$, A. Moewes$^1$}
\address{$^1$ Department of Physics and Engineering Physics, University of Saskatchewan, 116 Science Place, Saskatoon, Saskatchewan S7N 5E2, Canada}
\ead{john.mcleod@usask.ca}
\address{$^2$ Department of Physics, University of Osnabr\"{u}ck, Barbarastr. 7, D-49069 Osnabr\"{u}ck, Germany}
\address{$^3$ Institute of Metal Physics, Russian Academy of Sciences-Ural Division, 620990 Yekaterinburg, Russia}
\address{$^4$ Ames Laboratory U.S. DOE and Department of Physics and Astronomy, Iowa State University, Ames, Iowa 50011, USA}
\address{$^*$ Present address: Department of Chemistry, Princeton University, Princeton, New Jersey 08544, USA.}

\begin{abstract}
The electronic structure of BaFe$_2$As$_2$ doped with Co, Ni, and Cu has been studied by a variety of experimental and theoretical methods, but a clear picture of the dopant \textit{3d} states has not yet emerged. Herein we provide experimental evidence of the distribution of Co, Ni, and Cu \textit{3d} states in the valence band. We conclude that the Co and Ni \textit{3d} states provide additional free carriers to the Fermi level, while the Cu \textit{3d} states are found at the bottom of the valence band in a localized \textit{3d}$^{10}$ shell. These findings help shed light on why superconductivity can occur in BaFe$_2$As$_2$ doped with Co and Ni but not Cu.
\end{abstract}

\pacs{71.20.Dg, 74.25.Jb}
\submitto{\JPCM}

\section{Introduction}
Since the discovery of superconductivity in LaFeAsO, there has been a great deal of research in superconductivity in iron pnictides (see Reference \cite{feas_collection} and references therein). Superconductivity has subsequently been found in derivatives of the parent compounds \textit{A}Fe$_2$As$_2$ (where \textit{A} = Ca, Sr, Ba or Eu), Li(Na)FeAs, and FeSe. The basic features that are common to many of these parent compounds is the antiferromagnetic (AFM) ordering of the Fe spins at a N\'{e}el temperature T$_N \sim$ 100-200 K, and the quasi-2D nature of the electronic structure.

Materials based on BaFe$_2$As$_2$ (Ba122) are considered suitable model systems for the study of iron pnictide-based superconductors because they are structurally simple and large crystals can be grown.~\cite{ni_08,sefat_08} The crystal structure of Ba122 at room temperature is tetragonal, with a ThCr$_2$Si$_2$-type structure (space group \textit{I4/mmm}).~\cite{pfisterer_80} Upon cooling below $\sim140$ K, Ba122 undergoes a structural tetragonal-orthorhombic transition which is linked with spin-density wave formation. Like many iron pnictides superconductivity in Ba122 can be realized if the low temperature phase transition is suppressed via hydrostatic pressure or non-isovalent doping.~\cite{chu_09,canfield_09}

Atomic substitution within the Fe-As layer --- particularly on the Fe sites --- of Ba122 seems to be important for understanding of the origin of superconductivity in these compounds. In an itinerant model the substitution of a small amount of Fe by another \textit{3d} element is expected to be similar to indirect doping, since only the total count of electrons is significant, and a rigid-band picture should work in the first approximation. On the other hand, in a picture with localized \textit{3d} electrons, doping on the Fe site should directly affect the electron correlations in the Fe–As layers and a behavior quite different from the indirect doping should evolve. For instance, in cuprates the substitution of a few percent Ni or Zn on the Cu site leads to a strong reduction of T$_c$. In this context several groups recently investigated the properties of  Ba122:\textit{M} (where \textit{M} = Co, Ni, Cu, Zn) solid solutions.~\cite{mandrus_10,ni_10}

The properties of Ba122 doped with transition metals (Ba122:\textit{M}) was initially expected to follow the rigid-band model, and this approach worked well for Ba122:Co.~\cite{kemper_09,bittar_11} However the rigid-band model did not work well for Ba122:Ni, and Ba122:Cu fails to exhibit superconductivity all together.~\cite{canfield_09} The electronic structure of Ba122:\textit{M} has subsequently been examined in a variety of studies, and a somewhat confusing picture has emerged: DFT studies suggest that Co, Ni, and Cu are all isovalent with Fe,~\cite{wadati_10} while Hall measurements suggest an increase in free electron density,~\cite{rullier_albenque_09, butch_10, fang_09} and M\"{o}ssbauer measurements suggest no change in the localized \textit{d} shell due to doping.~\cite{khasanov_11}

A direct experimental study of the dopant electronic structure in Ba122:\textit{M} is necessary not only to verify the calculated electronic structure, but also to provide empirical estimates of the valency of the dopants. X-ray emission and absorption spectroscopy (XES and XAS) provide an element and symmetry specific probe of the occupied and unoccupied states, respectively, and therefore are excellent tools to study the electronic structure specifically related to the doping in Ba122:\textit{M}. We have measured the transition metal (Fe, Co, Ni, Cu) resonant and non-resonant \textit{L}$_{2,3}$ XES and XAS of single-crystal Ba122:\textit{M}, and complimented these measurements with core- and valence-level X-ray photoelectron spectroscopy (XPS). These experimental data are analyzed with the help of density functional theory (DFT) calculations of the electronic structure of Ba122:\textit{M}.

\section{Experimental and Theoretical Methods}

Single crystals of Ba122:\textit{M} (for doping concentrations of $\sim 0.07$ on Fe sites) with \textit{M} = Co, Ni, Cu were grown in a similar manner as described in Reference \cite{ni_08}. Actual doping levels (rather than nominal) were determined via wave-length dispersive spectroscopy (WDS) analysis using the electron probe microanalyzer of a JEOL JXA-8200 electron microprobe. Powder X-ray diffraction spectra with a Si standard were measured using a Rigaku MiniFlex and unit cell parameters were extracted using the ``UNITCELL'' analysis package. Electrical resistivity measurements were made using a standard 4-probe configuration and Quantum Design PPMS (Physical Property Measurement System) and MPMS (Magnetic Property Measurement System) units to provide the temperature/field environment. Pure Fe, Co, Ni, and Cu metals, and high-purity crystals of binary metal oxides (FeO, CoO, NiO, and CuO) were also obtained for reference measurements.

The resonant and non-resonant X-ray emission spectroscopy (XES) measurements of Ba122:\textit{M} were performed at the soft X-ray fluorescence endstation of Beamline 8.0.1 at the Advanced Light Source in the Lawrence Berkeley National Laboratory.~\cite{jia_95} The endstation uses a Rowland circle geometry X-ray spectrometer with spherical gratings and an area sensitive multichannel detector. We have measured the resonant and non-resonant metal (Fe, Co, Ni, Cu) \textit{L}$_{2,3}$ (probing the \textit{3d},\textit{4s} $\rightarrow$ \textit{2p} transition) XES spectra. The instrument resolving power (E/$\Delta$E) for the XES spectra was about 10$^{3}$. The X-ray absorption spectroscopy (XAS) measurements of the metal \textit{L}$_{2,3}$ edges in Ba122:\textit{M} were preformed at the spherical grating monochromator (SGM) beamline at the Canadian Light Source.~\cite{regier_07} The XAS measurements of Ba122:\textit{M} and pure metals were taken in the bulk-sensitive total fluorescence yield (TFY) mode to minimize surface oxidation effects,~\footnote{Our samples were cleaved under dry N$_2$ and directly transferred to vacuum, however \textit{in situ} vacuum cleaving capabilities were not available on the SGM beamline so some surface oxidation occurred; there was probably trace amounts of O$_2$ leaking into the N$_2$ environment.} while the XAS measurements of the metal oxides were taken in the bulk-sensitive and self absorption-free inverse partial fluorescence yield (IPFY) mode~\cite{achkar_11}. The instrumental resolving power for all XAS measurents was about 5$\times$10$^3$. The excitations for the resonantly excited XES measurements were determined from the XAS spectra; the chosen energies corresponded to the location of the \textit{2p}$_{3/2}$ and \textit{2p}$_{1/2}$ excitation thresholds, one energy between the two, and one energy well above resonance for both edges.~\footnote{XAS measurements were also performed at Beamline 8.0.1 just prior to the XES measurements, the resolving power and signal-to-noise ratio of these XAS measurements was poorer than that of the XAS measurements taken at the SGM beamline, and are not reported here. These measurements were used, however, to ensure the proper calibration between the XAS spectra from the SGM and the XES spectra from Beamline 8.0.1.}

\begin{figure}
\begin{center}
\includegraphics[width=3in]{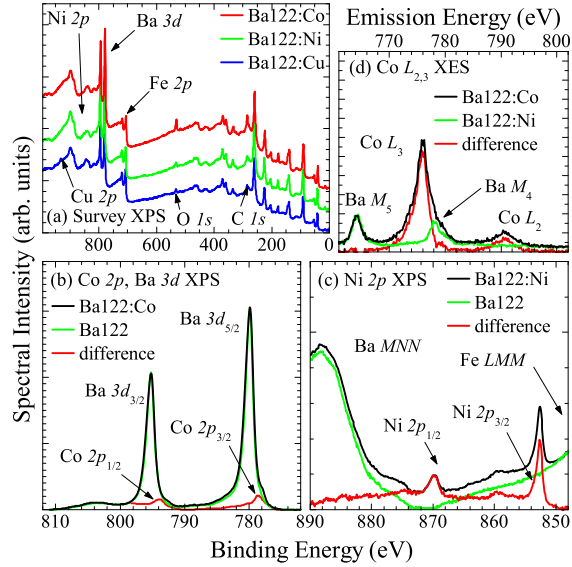}
\caption{Counter-clockwise from top left: (a) The survey XPS measurements for Ba122:\textit{M}. Note the relatively weak C, O \textit{1s} peaks indicating a high-quality, pure sample. (b) The Ba \textit{3d} and Co \textit{2p} core XPS spectra, showing how the Ba \textit{3d} peaks in the XPS spectrum of Ba122:Co were removed by subtracting the Ba \textit{3d} XPS spectrum of Ba122, leaving only the Co \textit{2p}-related features of Ba122:Co. (c) The Ba \textit{M}$_{4,5}$ and Co \textit{L}$_{2,3}$ XES spectra, showing how the Ba \textit{M}$_{4,5}$ emission in the XES spectrum of Ba122:Co by subtracting the Ba \textit{M}$_{4,5}$ XES emission from Ba122:Ni, leaving only the Co \textit{L}$_{2,3}$-related emission of Ba122:Co. (d) The Ni \textit{2p} XPS of Ba122:Ni, showing how the Ba \textit{MNN} and Fe \textit{LMM} Auger emission was subtracted using the spectrum from Ba122. (Color online.)}
\label{fig:survey}
\end{center}
\end{figure}

X-ray photoelectron spectroscopy (XPS) measurements were obtained using a Perkin-Elmer PHI 5600 ci Multitechnique System with monochromatized Al \textit{K}$_\alpha$ radiation (with a full width at half-maximum, FWHM, of 0.3 eV). The energy resolution of the spherical capacitor analyzer was adjusted to approximately $\Delta$E = 0.45 eV. The pressure in the ultra-high vacuum chamber was in the 10$^{-10}$ mbar range during the measurements. The Ba122:\textit{M} crystals were cleaved \textit{in situ}. The surface contamination was monitored by measuring the O \textit{1s} and C \textit{1s} core level spectra before and after our measurements, the relatively low intensity of these spectra indicate the high purity of our samples. The XPS survey scans are shown in Figure \ref{fig:survey}(a), for further details see our previous XPS measurements of CaFe$_2$As$_2$.~\cite{kurmaev_09}

The XES and XAS measurements were calibrated such that the \textit{L}$_3$ portion of the non-resonant XES in pure metals (Fe, Co, Ni, and Cu) matched the transition from valence XPS~\cite{fadley_68} to the \textit{2p}$_{3/2}$ core level. This places the peak of the measured \textit{L}$_3$ X-ray emission at 705.9 eV, 777.4 eV, 852.3 eV, and 930.2 eV for Fe, Co, Ni, and Cu, respectively. The XAS spectra were calibrated with respect to the XES spectra by matching the elastic scattering peaks in an XES spectrum to the corresponding excitation energy in the XAS spectrum.

Our analysis of the Co-doping in Ba122:Co is slightly complicated by the fact that the binding energy and the spin-orbit splitting of Ba \textit{3d} states is almost the same as that of the Co \textit{2p} states. Because Ba is present in a much larger concentration than Co, the Ba \textit{3d} contribution dominates the core XPS and XAS spectra in Ba122:Co. Fortunately the Ba \textit{3d} spectrum is relatively featureless in both cases. To isolate the Co \textit{2p} spectrum, we measured the Ba \textit{3d} core XPS for Ba122, and subtracted this spectrum from the Ba \textit{3d}/Co \textit{2p} XPS spectrum of Ba122:Co, as shown in Figure \ref{fig:survey}(b). In the case of the Ba \textit{M}$_{4,5}$/Co \textit{L}$_{2,3}$ XAS of Ba122:Co, the Co \textit{L}$_{3}$ absorption edge was sufficiently separated from the Ba \textit{M}$_{5}$ absorption edge, so no subtraction of the Ba \textit{M}$_{4,5}$ XAS was necessary. Since Ba has almost no occupied \textit{5p} or \textit{4f} states in the valence band, the Ba \textit{M}$_{4,5}$ XES was much lower intensity than the Co \textit{L}$_{2,3}$ XES in Ba122:Co, even though there was a much greater concentration of Ba than Co. Still, the Ba \textit{M}$_{4,5}$ contribution was subtracted from the Co \textit{L}$_{2,3}$ XES spectrum of Ba122:Co by using the Ba \textit{M}$_{4,5}$ XES spectrum of Ba122:Ni, as shown in Figure \ref{fig:survey}(c). In a related note, our analysis of the Ni \textit{2p} XPS in Ba122:Ni was slightly complicated by the relatively large intensity of Ba \textit{MNN} and Fe \textit{LMM} Auger emission that bracket the Ni \textit{2p} XPS spectrum when an Al \textit{K}$_\alpha$ source was used. As with the Ba \textit{3d} XPS in Ba122:Co, the Ba \textit{MNN} and Fe \textit{LMM} Auger contribution was subtracted from the Ni \textit{2p} XPS spectrum using the spectrum of Ba122, as shown in Figure \ref{fig:survey}(d).

All band structure calculations were performed using density functional theory (DFT) with the full-potential linearized augmented plane-wave (FP-LAPW) method as implemented in the WIEN2k code.~\cite{blaha_01} For the exchange-correlation potential we used the Perdew-Burke-Ernzerhof variant of the generalized gradient approximation (GGA).~\cite{perdew_96} The experimentally determined lattice parameters of ambient Ba122 ($a$ = 3.9635 \AA, $c$=13.022 \AA)~\cite{sefat_08} were used in our calculations. We replaced one of iron sites in a $2a \times 2a\times 2c$ supercell by a dopant (Co, Ni, or Cu) to account for the effect of doping. The resulting unit cell has a formula Ba\textit{M}$_{1/8}$Fe$_{15/8}$As$_2$, which is close to our experimental level of doping. The Brillouin zone was integrated over a grid of about 2500 total \textit{k}-points (the specific grid sizes depended on the doping sites). For the calculation the atomic sphere radii were chosen as R$_{Ba}$ = 2.5, R$_{Fe,Co,Ni,Cu}$ = 2.39 and R$_{As}$ = 2.12 a.u. These were selected so that the spheres are nearly touching. To investigate the sensitivity of dopant location with in the lattice, three different dopant sites were chosen, such that the symmetries of the $2a \times 2a\times 2c$ supercell were \textit{Pmmm}, \textit{P-4m2}, and \textit{I-4m2}. The differences in the density of states (DOS) for these different structures were minor, and certainly not resolvable by our experimental methods. We therefore simply averaged the DOS for each of these structures together, and we consider the result sufficient to represent the presumably random location of the dopants within the Ba122 lattice. XES and XAS spectra were calculated based on the DFT electronic structure using the ``XSPEC'' package.~\cite{schwarz_79} These spectra were Lorentzian and then Gaussian broadened to account for lifetime and instrumental broadening. For the former broadening the FWHM was energy-dependent as described in Reference \cite{schwarz_79} with $\Gamma_0 = 0.1$ eV, $W = 0.8$ eV, while for the latter the resolving power of the equipment was used to determine the FWHM.

\section{Results and Discussion}

There are two basic questions regarding the electronic structure of Ba122:\textit{M} that we hope to answer, namely:
\begin{enumerate}
\item What is the valency of the \textit{3d} transition metal dopants in Ba122:\textit{M}?
\item Where do the occupied \textit{3d} dopant states reside in the valence band of Ba122:\textit{M}?
\end{enumerate}
Since X-ray spectroscopy provides an element and symmetry specific probe of the electronic structure, it is an ideal tool to use in answering these questions. We will proceed by analyzing the spectra of elements which are common between all Ba122:\textit{M} samples studied herein, in particular the Fe \textit{2p} XPS and Fe \textit{L}$_{2,3}$ XES and XAS. We will then focus on the \textit{2p} XPS and \textit{L}$_{2,3}$ XES and XAS of the dopants in Ba122:\textit{M} compared to those of pure metals and metal oxides. Finally our valence XPS and \textit{L}$_{2,3}$ XES and XAS spectra provide an experimental map of the valence and conduction bands, and these can be compared directly to the calculated electronic structure.

\subsection{Core-Level Spectroscopy of the Ba122:\textit{M} Lattice}

\begin{figure}
\begin{center}
\includegraphics[width=3in]{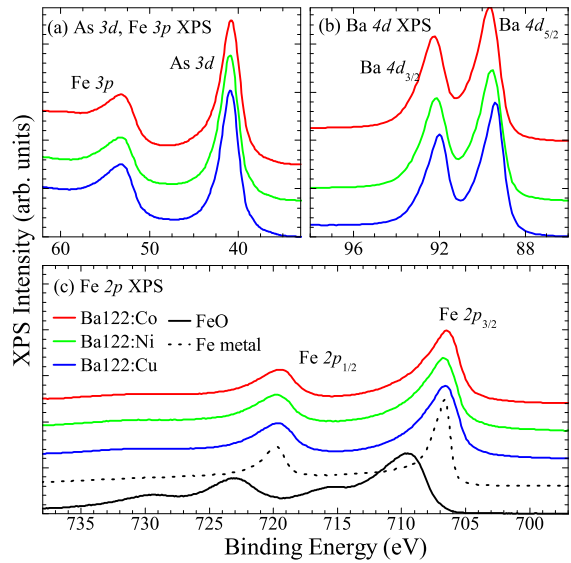}
\caption{Some XPS core spectra common to all of our Ba122:\textit{M} samples. (a) The Fe \textit{3p} and As \textit{3d} XPS spectra. (b) The Ba \textit{4d}$_{3/2}$ and \textit{4d}$_{5/2}$ XPS spectra. (c) The Fe \textit{2p}$_{1/2}$ and Fe \textit{2p}$_{3/2}$ XPS spectra, as well as reference Fe \textit{2p} XPS spectra for comparison. Note the lack of satellites or shoulders in all of the Ba122:\textit{M} XPS spectra shown here. (Color online.)}
\label{fig:core_xps}
\end{center}
\end{figure}

To start with, we note that the XPS core spectra of As \textit{3d} and Ba \textit{4d} in all of our samples are quite featureless, as shown in Figure \ref{fig:core_xps}(a,b), and have no shoulders, contrary to previously published spectra~\cite{jong_09}. These shoulders are usually attributed to surface contributions and help confirm the high quality of our samples. There is also no peak splitting in the As \textit{3d} XPS spectra as was found in LaFeAsO$_{0.9}$F$_{0.1}$, this splitting was attributed to a strong As \textit{4p}---Fe \textit{3d} hybridization.~\cite{garcia_08} However a two-layered iron pnictide like Ba122 should have a similar amount of hybridization as a one-layered iron pnictide like LaFeAsO, so the absence of this feature in our spectra suggests that it is not due to chemical bonding.

The Fe \textit{2p} XPS spectra of all Ba122:\textit{M} samples shown in Figure \ref{fig:core_xps}(c) provide further evidence that these systems are weakly or at most moderately correlated. The Fe \textit{2p} XPS spectra for Ba122:\textit{M} are all quite similar to the spectrum of pure iron,~\cite{moulder_92} and show none of the charge-transfer satellites and chemical shifts inherent to iron oxides.~\cite{galakhov_97} These spectra preclude the possibility that Ba122:\textit{M} is a strongly correlated material. Indeed, these spectra indicate that the iron sites are rather metal-like in nature, and suggest that the iron may be closer to a formal valency of 0 rather than 2+.

\begin{figure}
\begin{center}
\includegraphics[width=3in]{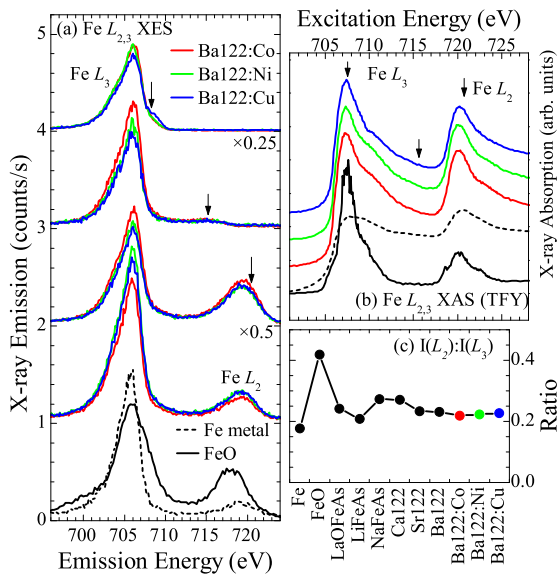}
\caption{The Fe \textit{L}$_{2,3}$ XES and XAS for Ba122:\textit{M}. (a) The resonant and non-resonant Fe \textit{L}$_{2,3}$ XES for Ba122:\textit{M}, the arrows indicate the excitation energy. The bottom set of spectra are excited at 730 eV, far from resonance. The non-resonant Fe \textit{L}$_{2,3}$ XES spectrum of Fe and FeO are plotted for reference. (b) The Fe \textit{L}$_{2,3}$ XAS spectra of Ba122:\textit{M} plotted in bulk-sensitive TFY mode. The Fe \textit{l}$_{2,3}$ XAS spectra of Fe metal and FeO are plotted for reference, note the strong self-absorption present in the Fe metal spectrum. (c) A comparison of the ratio of the intensity of the Fe \textit{L}$_2$ and \textit{L}$_3$ X-ray emission, when excited far from resonance, for Ba122:\textit{M}, Fe, FeO, and several other iron pnictides.~\cite{kurmaev_09,kurmaev_09_2,yang_09} (Color online.)}
\label{fig:fe_xs}
\end{center}
\end{figure}

This description of the Fe sites is supported by our resonant XES spectra. The Fe \textit{L}$_{2,3}$ XES spectra of Ba122:\textit{M} for both resonant and non-resonant excitation are also quite featureless and again quite similar to the corresponding spectrum of Fe metal, as shown in Figure \ref{fig:fe_xs}(a). The resonant \textit{L}$_2$ and \textit{L}$_3$ spectra of Ba122:\textit{M} show the same features as the non-resonant spectrum; there are no additional features due to multiplet effects or other inelastic scattering. This is consistent with resonant Fe \textit{L}$_{2,3}$ XES measurements of other iron pnictides,~\cite{kurmaev_09, kurmaev_09_2,yang_09} and unlike resonant Fe \textit{L}$_{2,3}$ XES measurements of FeO.~\cite{prince_05} The lack of discernable difference between the Fe \textit{L}$_{2,3}$ XES spectra of Ba122:\textit{M} and Fe metal indicates there is not a significant amount of metal-metal charge transfer (between Fe and the \textit{M} dopants) in Ba122:\textit{M}.

The Fe \textit{L}$_{2,3}$ XAS spectra of Ba122:\textit{M} are also rather featureless, and again similar to the spectrum for Fe metal, as shown in Figure \ref{fig:fe_xs}(b). Note that the higher Fe density in Fe metal means there is more self-absorption in this spectrum and hence the lower amplitude of the spectral peaks, however note that an undistorted Fe \textit{L}$_{2,3}$ XAS spectrum has practically no fine structure.~\cite{kruger_09} These spectra are essentially the same as the Fe \textit{L}$_{2,3}$ XAS spectra of iron pnictides previously reported.~\cite{yang_09,kroll_08,bondino_09} The lack of many multiplet satellites from a \textit{2p}$\rightarrow$\textit{3d}$\bar{L}$ transition in these spectra again suggests a mostly metallic environment. 

Based on the filling of \textit{2p} core holes, one would naively expect the ratio of the intensity of the \textit{L}$_2$ and \textit{L}$_3$ X-ray emission lines (I(\textit{L}$_2$)/I(\textit{L}$_3$)) to equal the ratio of the occupancy of the \textit{2p}$_{1/2}$ and \textit{2p}$_{3/2}$ levels --- namely I(\textit{L}$_2$)/I(\textit{L}$_3$) = 0.5. However in solids radiationless Coster-Kronig \textit{L}$_2$\textit{L}$_3$\textit{M}$_{4,5}$ transitions can reduce the intensity of the \textit{L}$_2$ band, particularly in metallic systems.~\cite{krause_79,kurmaev_05} A non-resonant X-ray emission spectrum and its I(\textit{L}$_2$)/I(\textit{L}$_3$) intensity ratio can therefore provide a qualitative spectrum-based estimate of the number of free carriers. For Ba122:\textit{M}, I(\textit{L}$_2$)/I(\textit{L}$_3$) $\approx 0.22$, a little higher than in Fe metal where I(\textit{L}$_2$)/I(\textit{L}$_3$) $\approx 0.18$, but much lower than in correlated FeO where I(\textit{L}$_2$)/I(\textit{L}$_3$) $\approx 0.42$, as shown in Figure \ref{fig:fe_xs}(c). These spectral observations are consistent with previous studies suggesting that the \textit{3d} states in iron pnictides are weakly or at most moderately correlated.~\cite{kurmaev_09,kurmaev_09_2,kurmaev_08,anisimov_09}

\subsection{Core-level Spectroscopy of Dopants in Ba122:\textit{M}}

\begin{figure}
\begin{center}
\includegraphics[width=3in]{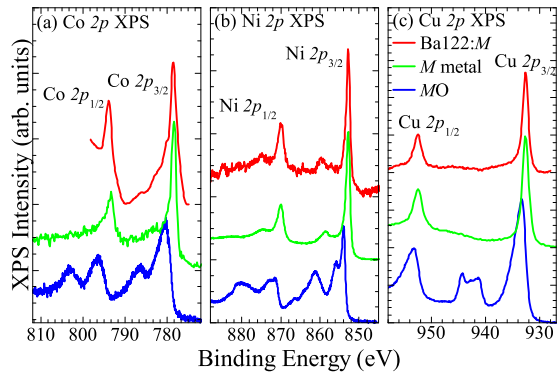}
\caption{The doped \textit{2p} XPS spectra for Ba122:\textit{M}, and their respective pure metal and binary oxide counterparts. (a) The Co \textit{2p} XPS spectra for CoO, Co metal, and Ba122:Co. Note that, as discussed in the previous section, the Ba \textit{3d} contribution was subtracted from the Ba122:Co spectrum. (b) The Ni \textit{2p} XPS spectra for NiO, Ni metal, and Ba122:Ni. (c) The analogous spectra for CuO, Cu metal, and Ba122:Cu. (Color online.)}
\label{fig:dope_xps}
\end{center}
\end{figure}

To focus now on the effect of the Co, Ni, and Cu dopants, we turn to the XPS core spectra of these elements. Just as for the Fe \textit{2p} XPS spectra in Ba122:\textit{M}, the \textit{2p} XPS spectra of the doping elements are fairly featureless and almost identical to the \textit{2p} XPS spectra of the corresponding pure transition metals, as shown in Figure \ref{fig:dope_xps}. The satellite-rich \textit{2p} XPS spectra of the corresponding binary transition metal oxides, on the other hand, bears very little similarity in both shape and energy shift to the spectra of Ba122:\textit{M}. This suggests that the doped transition metals are just as metallic as the Fe in Ba122:\textit{M}, and again that they may have a formal valency closer to 0 than 2+.

\begin{figure}
\begin{center}
\includegraphics[width=3in]{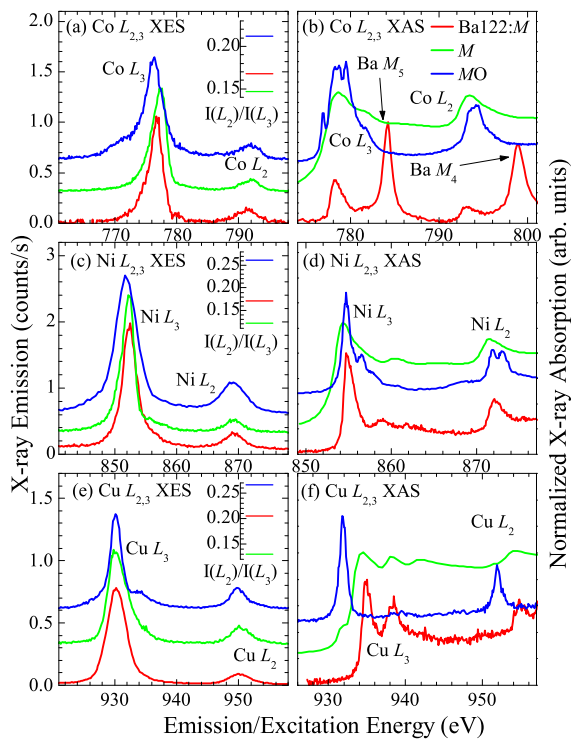}
\caption{The doped \textit{L}$_{2,3}$ XES and XAS spectra for Ba122:\textit{M}, and their respective pure metal and binary oxide counterparts. (a) The Co \textit{L}$_{2,3}$ XES and (b) XAS spectra for CoO, Co metal, and Ba122:Co. The inset in (a) shows the I(\textit{L}$_2$)/I(\textit{L}$_3$) ratio for the Co \textit{L}$_{2,3}$ XES of these materials. Note that, as discussed in the previous section, the Ba \textit{M}$_{4,5}$ contribution was subtracted from the Ba122:Co XES spectrum, the Ba \textit{3d} contribution is still visible in the Ba122:Co XAS spectrum. (b) The Ni \textit{L}$_{2,3}$ XES, I(\textit{L}$_2$)/I(\textit{L}$_3$) ratio (inset), and (c) the Ni \textit{L}$_{2,3}$ XAS spectra for NiO, Ni metal, and Ba122:Ni. (e) and (f), the analogous spectra for CuO, Cu metal, and Ba122:Cu. (Color online.)}
\label{fig:dope_xs}
\end{center}
\end{figure}

The dopant \textit{L}$_{2,3}$ XES and XAS for Ba122:Co and Ba122:Ni tell the same story as the Fe X-ray spectra: in each case the metal \textit{L}$_{2,3}$ XES spectrum is quite similar in shape to that of the pure metal, rather than the binary oxide, as shown in Figure \ref{fig:dope_xs}(a-d). The CoO and NiO low energy satellites at $\sim$769 and $\sim$846 eV, respectively, are not seen in the spectra of Ba122:(Co,Ni). Likewise, the dopant \textit{L}$_{2,3}$ XAS spectra are similar in shape to the \textit{L}$_{2,3}$ XAS of the pure metal (see Figure \ref{fig:dope_xs}(b,d)),~\footnote{Note that the bulk sensitive TFY XAS of the dopants has much less self-absorption (since the doping concentration is relatively low) than the pure metals. Note also that the metal oxide XAS was measured in IPFY mode, this makes these spectra self-absorption free. IPFY mode would be ideal for Ba122:\textit{M} as well, but these samples lack a sharp absorption edge in the soft X-ray region but below the Fe \textit{L}$_{2,3}$ edge, so there is no absorption channel for ``inversion''.} again suggesting at most a weak metal-ligand interaction. These observations suggest that the dopants are closer to Co$^0$, Ni$^0$ metals with free \textit{3d} electrons than Co$^{2+}$, Ni$^{2+}$ atoms in a strong ligand bonding environment.

The situation for Ba122:Cu is slightly different: here the Cu I(\textit{L}$_2$)/I(\textit{L}$_3$) for Ba122:Cu is closer to CuO than Cu metal (note the intensities of the \textit{L}$_2$ and \textit{L}$_3$ X-ray emission in Figure \ref{fig:dope_xs}(e)), however the Cu \textit{L}$_{2,3}$ XAS for Ba122:Cu is clearly quite similar to that of Cu metal. This latter comparison is crucial because if there were \textit{3d} holes (as in the case of CuO, a \textit{d}$^9$ system) or a strong ligand interaction hybridizing with \textit{3d} states (as is the case for Cu$_2$O), the XAS for Ba122:Cu would show a strong pre-edge peak (as is seen in the Cu \textit{L}$_{2,3}$ XAS of CuO in Figure \ref{fig:dope_xs}(f)) from unoccupied \textit{3d} states.~\cite{grioni_92} These two measurements suggest the following picture: the Cu in Ba122:Cu is a \textit{3d}$^{10}$ system (likely Cu$^0$, although Cu$^{1+}$ cannot be ruled out), however the Coster-Kronig transitions are somewhat suppressed in Ba122:Cu compared to Cu metal (note that by referring back to Figure \ref{fig:dope_xps}(c) we see that the ratio of the Cu \textit{2p}$_{1/2}$ and Cu \textit{2p}$_{3/2}$ XPS spectral intensities is roughly the same in Ba122:Cu and Cu metal, indicating that ratio of \textit{2p}$_{1/2}$ and \textit{2p}$_{3/2}$ core holes is the same for both materials). The suppression of Coster-Kronig transitions suggests the Cu \textit{3d} states are localized.

\subsection{Valence Electronic Structure in Ba122:\textit{M}}

\begin{figure}
\begin{center}
\includegraphics[width=3in]{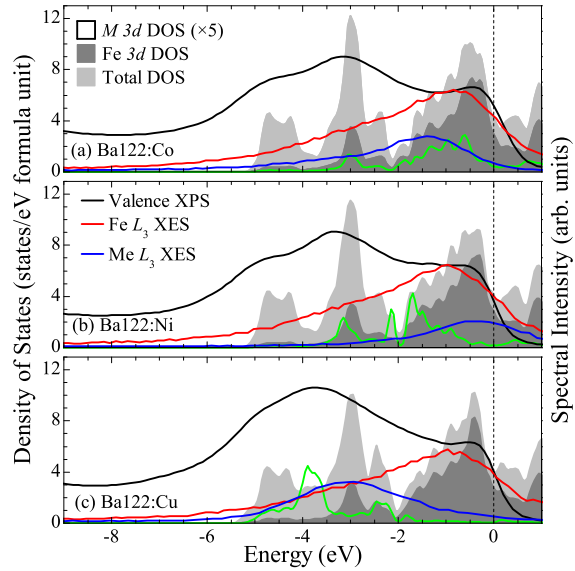}
\caption{The calculated DOS for Ba122:\textit{M}. The total DOS, Fe \textit{3d} DOS, and the dopant \textit{3d} DOS is plotted for (a) Ba122:Co, (b) Ba122:Ni, and (c) Ba122:Cu. The Fermi level (E$_F$) is at 0 eV. The measured valence XPS spectrum for each material is superimposed on the calculated DOS. Finally, using the core binding energies obtained from our XPS measurements, the non-resonant metal \textit{L}$_3$ XES spectra are superimposed on the calculated DOS, these represent an experimental estimation of the distribution and energy of the partial DOS. The measured spectra have been arbitrarily scaled to approximately match the intensity of the calculated DOS. (Color online.)}
\label{fig:dos}
\end{center}
\end{figure}

Our calculated DOS for Ba122:\textit{M} is essentially the same as those previously reported for doped and undoped Ba122 (note our DOS spectra are fairly broad because we averaged the DOS of several doped structures together).~\cite{sefat_08,wadati_10,shein_08,mandrus_10} The calculated DOS is also in good agreement with our measured valence XPS spectra, as shown in Figure \ref{fig:dos}. As previously shown,~\cite{kurmaev_09} the valence region may be divided into three basic bands: the first, from 0 to -2 eV is mainly Fe \textit{3d} states, the second band from -2 to -4 eV is hybridized Fe \textit{3d} --- As \textit{4p} states, and the third and weakest band, from -4 eV to -6 eV, is mostly As \textit{4p}. The \textit{3d} states of the Co, Ni, and Cu dopants is centred progressively deeper in the valence band with increasing dopant atomic number.~\cite{wadati_10} Using the binding energies obtained from our core level XPS measurements, we can shift the transition metal \textit{L}$_3$ XES spectra to an energy scale that is consistent with the valence XPS measurements (the \textit{L}$_2$ XES spectrum could also be used, but it is less intense and has a poorer signal-to-noise ratio than the \textit{L}$_3$ XES spectrum). We therefore have a purely experimental probe of not only the total DOS (from XPS), but also the local partial DOS of \textit{3d}-symmetry of the transition metals.

This approach shows that the Fe \textit{3d} states involved in the Fe \textit{L}$_3$ XES spectrum coincide with the the sharp peak at the Fermi level in the measured XPS for all Ba122:\textit{M}, as we might expect. The peak in the Fe \textit{L}$_3$ XES does not agree exactly with the XPS peak due to the inherent broadening in the XES spectrum. Likewise, the Co and Cu states agree reasonably well with the calculated partial DOS. The Ni \textit{3d} states, however, appear to be about 1 eV higher in energy in the measured Ni \textit{L}$_3$ XES than suggested by the calculation. This shift is greater than the experimental resolution, but the real culprit here may be the absolute energy scale of our XES spectra. While we can be fairly confident that the \textit{2p}$_{3/2}$ binding energy and \textit{L}$_3$ X-ray emission energy of the transition metals in Ba122:\textit{M} is quite close to that of the equivalent pure metals (refer back to Figures \ref{fig:dope_xps} and \ref{fig:dope_xs}), we can not necessarily assume that the transition metals in Ba122:\textit{M} will have the same structure near the Fermi level as the equivalent pure metals. In short, while XPS can provide accurate binding energies, it is difficult to provide an absolute energy scale for the XES spectra (at least if one wants greater accuracy than $\pm 1$ eV) because the Fermi level often can not be accurately determined.

\begin{figure}
\begin{center}
\includegraphics[width=3in]{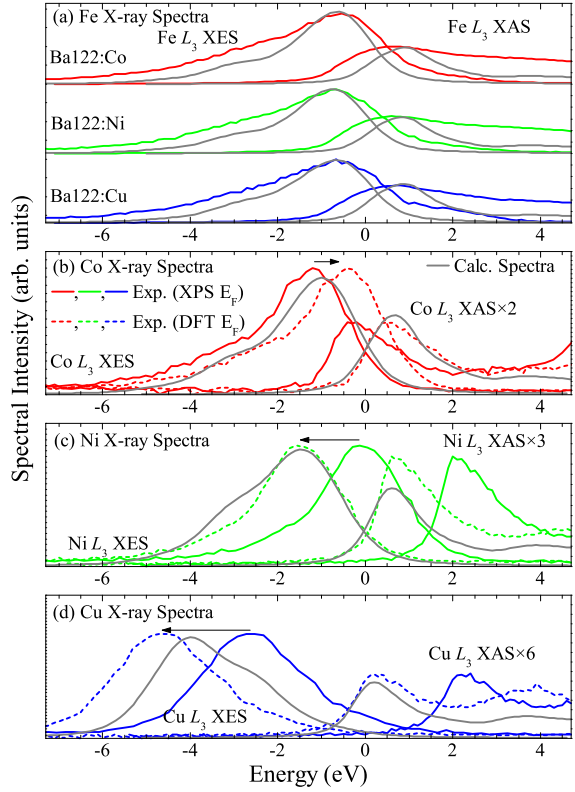}
\caption{The calculated and measured metal \textit{L}$_3$ XES and XAS spectra for Ba122:\textit{M}. (a) The Fe spectra in all Ba122:\textit{M}, and the dopant spectra in (b) Ba122:Co, (c) Ba122:Ni, and (d) Ba122:Cu. The Fermi level (E$_F$) is at 0 eV. The measured valence XPS spectrum for each material is superimposed on the calculated DOS. Finally, using the core binding energies obtained from our XPS measurements, the non-resonant metal \textit{L}$_3$ XES spectra are superimposed on the calculated DOS, these represent an experimental estimation of the distribution and energy of the partial DOS. The measured spectra have been arbitrarily scaled to approximately match the intensity of the calculated DOS. (Color online.)}
\label{fig:calc_xs}
\end{center}
\end{figure}

Perhaps a better method of comparing the measured XES and XAS spectra to the calculated partial density of states is to align the calculated and measured spectra directly. Since Ba122:\textit{M} is a bad metal with no band gap, the onset of the XAS spectra should be quite close to the Fermi level. Even though the \textit{2p} core hole may distort the shape of the experimental XAS spectrum, it should not be able to shift the XAS spectrum below the Fermi level, especially if the material is not strongly correlated. Since the relative calibration between the measured XES and XAS spectra is quite accurate,~\footnote{For an instrumental resolving power $R$ and an excitation energy $E$, the relative calibration between XES and XAS is $\sim RE / \sqrt{N}$ where $N$ is the number of counts in the elastic scatter observed on the XES spectrometer, for these measurements $N \sim 30$ so the calibration is accurate to $0.1 - 0.2$ eV.} we can align the measured XAS with the calculated XAS and obtain the energy of the corresponding measured XES from that. The calculated and measured transition metal \textit{L}$_{3}$ XES and XAS for Ba122:\textit{M} is shown in Figure \ref{fig:calc_xs}.

The shape of both the calculated XES and XAS spectra is quite close to the corresponding measured spectra. The iron peak splitting (which can be more clearly seen in the Fe \textit{2p} XAS in Figure \ref{fig:fe_xs}) is not reproduced in the calculated Fe \textit{L}$_{3}$ XAS in Figure \ref{fig:calc_xs}(a), but because this is due to the Fe \textit{2p} core hole we do not expect it to show up in the XAS spectrum calculated from the ground state DOS. It is also difficult to compare the measured and calculated Co \textit{L}$_{3}$ XAS due to the distortion caused by the Ba \textit{M}$_{5}$ contribution to the measured spectrum shown in Figure \ref{fig:calc_xs}(b). Finally the calculated Ni \textit{L}$_{3}$ XES spectrum has a higher shoulder than the measured spectrum (at -3 eV in the calculated spectrum in Figure \ref{fig:calc_xs}(c)). Apart from these aspects, the shape of the remaining calculated spectra are quite close to the shape of the measured spectra.

In Figure \ref{fig:calc_xs} the measured spectra are plotted on the DOS energy scale in two ways: the solid colored lines indicate the spectra aligned using the XPS binding energy method discussed above (the ``XPS E$_F$'' method), just as these spectra are plotted in Figure \ref{fig:dos}. The dashed colored lines indicate the same measured spectra shifted so that the energy of the peak in the metal \textit{L}$_3$ XAS spectrum matches the one in the calculated spectrum (the ``DFT E$_F$'' method). As mentioned, this was done because we expect the absorption spectrum to be a better indicator of the Fermi level in bad metals like Ba122:\textit{M} than the emission spectrum, especially because for Ni and Cu the calculation suggests that the bulk of the occupied states are buried in the valence band, with very few occupied states at the Fermi level. It is clear that the XPS E$_F$ and the DFT E$_F$ both give essentially the same result for the Fe \textit{L}$_3$ XES and XAS, shown in Figure \ref{fig:calc_xs}(a). However because the Fe \textit{3d} states dominate the top of the valence band, this agreement is not unexpected.

There is reasonable agreement between the measured and calculated Co spectra in Figure \ref{fig:calc_xs}(b), although be examining the XAS we may be justified in shifting the measured Co spectra by as much as +0.5 eV. This shift would move the bulk of the Co \textit{3d} valence states, as probed by the measured Co \textit{L}$_3$ XES, by about 0.3 eV nearer to the Fermi level than predicted by DFT. This would put the bulk of the Co \textit{3d} states at the same energy as the bulk of the Fe \textit{3d} states, however we must point out that these energy shifts are fairly small compared to the instrumental resolution.

The energy of the measured Ni spectra in Figure \ref{fig:calc_xs}(c) aligned with the XPS E$_F$ method is about 1.3 eV higher than the calculated spectra. This difference is well within the experimental resolving power. If the peak in the measured Ni \textit{L}$_{3}$ XAS spectrum is aligned with the peak in the calculated spectrum (the DFT E$_F$ method), however, the measured Ni \textit{L}$_3$ XES spectrum is in almost perfect alignment with the calculated spectrum. This good agreement suggests that our absolute energy calibration for X-ray emission at the Ni \textit{L}$_3$ edge is not as accurate as the energy calibration for the Co or Fe \textit{L}$_3$ edges.

Finally, the XPS E$_F$ alignment of the Cu spectra in Figure \ref{fig:calc_xs}(d) shows an even larger discrepancy of about 2.0 eV compared to the calculated spectra. If the DFT E$_F$ alignment is used, the measured Cu \textit{L}$_3$ XES is now about 0.7 eV deeper in the valence band than the calculated spectrum.

\subsection{The Impact of \textit{3d} Dopants on Electronic Structure}

Ultimately, while the ``most valid'' method for aligning experimental spectra with calculations can be disputed, it is not possible to align both the XES and XAS with the calculated spectra for the Cu \textit{L}$_3$ XES and XAS in a manner which is consistent with expected energy shifts due to the core hole in the experimental XAS spectrum. It also seems that the predicted separation between the Co \textit{L}$_3$ XES and XASspectra does not agree with the experimental separation, although the discrepancy for Co is considerably smaller than the discrepancy for Cu. 

\begin{figure}
\begin{center}
\includegraphics[width=3in]{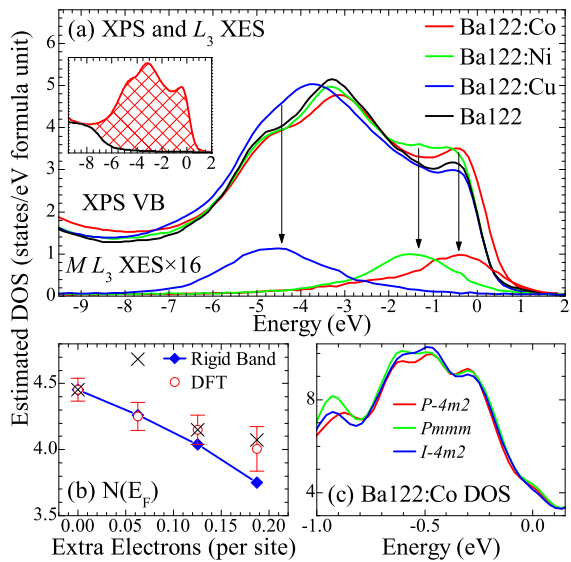}
\caption{The measured valence XPS and dopant \textit{L}$_3$ XES spectra, and behaviour at the Fermi level for Ba122:\textit{M}. (a) The total and partial occupied DOS estimated using the measured XPS and dopant \textit{L}$_{3}$ XES spectra (each XES spectrum was aligned using the DFT E$_F$ method). Each XPS (XES) spectrum has been scaled such that the integral of the background-subtracted spectrum from -8 to 0 eV matches the total number of valence (\textit{3d}) electrons per formula unit (with 0.07 doping). The inset shows an example of the XPS background subtraction in Ba122:Co; the background was approximated using an arctangent and the shaded area represents the region that was scaled to match the number of valence electrons. The XES background subtraction was just a straight line representing the noise threshold. (b) The number of states at the Fermi level, N(E$_F$), for the rigid band model, the supercell DFT calculation (the average for the three different doping structures, the error bars represent the standard deviation of this average). (c) The calculated total DOS for Ba122:Co near the Fermi level, showing the sensitivity of N(E$_F$) to the doping structure. (Color online.)}
\label{fig:fermi_level}
\end{center}
\end{figure}

The validity of the calculated \textit{L}$_{3}$ XAS to align the experimental spectrum is supported by comparing the measured valence XPS to the new alignment of the Co, Ni, and Cu \textit{L}$_3$ XES. In each case the relative intensity of the valence XPS of Ba122:\textit{M} is slightly enhanced in the region corresponding to the \textit{M} \textit{3d} states (estimated from the new alignment of the \textit{M} \textit{L}$_3$ XES), as shown in Figure \ref{fig:fermi_level}(a). This approach therefore suggests that the \textit{3d} states of Co in Ba122:Co hybridize almost entirely with the Fe \textit{3d} states - recall that the calculated DOS suggest that the bulk of the Co \textit{3d} states are about 0.3 eV lower than the Fe \textit{3d} states (see Figure \ref{fig:dos} and Reference \cite{wadati_10}). The Ni \textit{3d} states in Ba122:Ni are slightly less hybridized with the Fe \textit{3d} states, since the bulk of these states are between the Fe \textit{3d} and As \textit{4p} states, in agreement with the DFT predictions. Finally, the Cu \textit{3d} states in Ba122:Cu are quite weakly hybridized with the Fe \textit{3d} states, since the measured spectra suggest they form a filled \textit{3d} shell at the bottom of the valence band, about 0.7 eV lower than predicted by DFT.

Because of the overlap of Fe \textit{3d} and Co \textit{3d} states, we can attribute the success of the rigid band model to the fact that the Co \textit{3d} distribution is essentially the same as the Fe \textit{3d} distribution with a higher occupancy. In this regard it is tempting to compare the number of states at the Fermi level, N(E$_F$), calculated by our supercell DFT approach with the rough value predicted from the rigid band model.~\footnote{We estimate N(E$_F$) in the rigid band model by using the DOS of undoped Ba122 and finding the ``new'' Fermi level by extending the occupied states by 1/16, 1/8, and 3/16 electrons per Fe site for Co-, Ni-, and Cu-doping, respectively.} N(E$_F$)s for the rigid band model and the supercell DFT calculations are shown in Figure \ref{fig:fermi_level}(b). As we expect, the rigid band model estimates roughly a linear decrease in N(E$_F$) with an increase in extra electrons. The supercell approach provides a similar picture, giving essentially the same N(E$_F$) for supercell Ba122:Co as for the rigid band case. The supercell Ba122:Ni and Ba122:Cu, however, both have larger N(E$_F$) than the rigid band model predicts. We should caution, however, that the supercell approach can have significant variability in N(E$_F$) depending on the particular structure chosen for the doping. As mentioned above, and shown in Figure \ref{fig:fermi_level}(c), the basic DOS is the same for each doping level but for narrow energy ranges (such as E$_F \pm k_{B}T$) there can be significant differences. The error bars in Figure \ref{fig:fermi_level}(b) reflect the standard deviation in N(E$_F$) for the three different doping structures calculated here, they show that if only one structure for doping is considered one could possibly conclude that Ba122:Co, Ba122:Ni, and even Ba122:Cu agree with the rigid band model for N(E$_F$), we therefore caution against attributing too much meaning to fine structure in doped iron pnictide DFT calculations without first considering a broad range of structures. Since our experimental study suggests that the Co and Cu \textit{3d} states are 0.3 eV closer to, and 0.7 eV farther from, the Fermi level, respectively, than supercell DFT calculations suggest, we have reason to suppose that the actual N(E$_F$) is larger for Ba122:Co and smaller for Ba122:Cu than shown here. This theory is supported by the valence XPS measurements, as shown in Figure \ref{fig:fermi_level}(a) the XPS intensity near E$_F$ is highest for Ba122:Co (even higher than undoped Ba122) and lowest for Ba122:Cu (although admittedly this analysis relies on the accuracy of our normalization scheme), and a study on PrFeAsO$_{1-\delta}$ which found that reducing the number of oxygen vacancies increased the number Fe states at the Fermi level.~\cite{freelon_10}

Our experimental study therefore provides a separation between the electronic structure of the dopants that assist superconductivity (Co, Ni) and the dopants that suppress it (Cu). We suggest that all dopants have a formal valency of 0, and the Co --- and to a lesser extent the Ni --- \textit{3d} states are strongly hybridized with the Fe \textit{3d} states near the Fermi level. On the other hand, Cu has a \textit{3d}$^{10}$ shell that is located at the bottom of the valence band. The similarity between I(\textit{L}$_2$)/I(\textit{L}$_3$) in the Co, Ni \textit{L}$_{2,3}$ XES for Ba122:Co,Ni and pure Co, Ni metal suggests that the \textit{3d} states of these dopants are rather metallic (note that since the Ba \textit{M}$_4$ XES spectrum was subtracted from the Co \textit{L}$_3$ XES spectrum in Ba122:Co it is quite possible that I(\textit{L}$_2$)/I(\textit{L}$_3$) in this case should be even lower), while for Cu the higher I(\textit{L}$_2$)/I(\textit{L}$_3$) in the Cu \textit{L}$_{2,3}$ XES for Ba122:Cu suggests that the Cu \textit{3d} states are more localized in Ba122:Cu than in Cu metal. Our DFT calculation overestimates the energy of the Cu \textit{3d} states, this can be attributed to two reasons: the DFT calculation underestimates the \textit{d}-\textit{p} Coulomb repulsion on the Cu \textit{3d}$^{10}$ shell, and the Cu-dopants distort the local structure of As ligands (the DOS of Ba122 is quite sensitive to the As \textit{z} coordinate~\cite{singh_08}).

We therefore propose the following picture: Both the Co and Ni \textit{3d} states are free carriers and increase the electron density near the Fermi level, as suggested by recent Hall measurements.~\cite{rullier_albenque_09,butch_10,fang_09} These states are quite metallic and are not very localized, explaining why there is no observed change in electron density near Fe sites in M\"{o}ssbauer spectroscopy of Ba122:Co,Ni.~\cite{khasanov_11} The Co \textit{3d} states are located right at the Fermi level, coincident with the Fe \textit{3d} states. The extra \textit{3d} states from Co at the Fermi level may drive the reduction in the hole pocket observed by angle-resolved photoelectron spectroscopy measurements.~\cite{sekiba_09} On the other hand the concentration of Ni \textit{3d} states somewhat below the Fermi level may help sustain the hole pocket, promoting the hole conduction observed in Hall effect measurements.~\cite{olariu_11} The Cu \textit{3d} states, in contrast, form a localized \textit{3d}$^{10}$ shell at the bottom of the valence band. While Cu-doping will disrupt the crystal structure sufficiently to prevent the low temperature phase transition necessary for superconductivity to be realized,~\cite{canfield_09} the Cu dopants actually reduce the number of free carriers at the Fermi level.

\section*{Acknowledgments}
We gratefully acknowledge support from the Natural Sciences and Engineering Research Council of Canada (NSERC) and the Canada Research Chair program. This work was done with partial support of the Russian Science Foundation for Basic Research (Project No. 11-02-00022). Work at the Ames Laboratory was supported by the Department of Energy, Basic Energy Sciences under Contract No. DE-AC02-07CH11358. The Advanced Light Source is supported by the Director, Office of Science, Office of Basic Energy Sciences, of the U. S. Department of Energy under Contract No. DE-AC02-05CH11231. The Canadian Light Source is supported by NSERC, the National Research Council (NSC) Canada, the Canadian Institute of Health Research (CIHR), the Province of Saskatchewan, Western Economic Diversification Canada, and the University of Saskatchewan. The computational part of this research was enabled by the use of computing resources provided by WestGrid and Compute/Calcul Canada.

\section*{References}
\bibliographystyle{unsrt}
\bibliography{ba122_jpcm}

\begin{thebibliography}{10}

\bibitem{feas_collection}
{Special issues of Physica C {\textbf{468}} 313-674 (2009) and New J. Phys.
  {\textbf{11}} 023001-024023 (2009) on superconducting pnictides.}

\bibitem{ni_08}
N.~Ni, S.~L. Bud'ko, A.~Kreyssig, S.~Nandi, G.~E. Rustan, A.~I. Goldman,
  S.~Gupta, J.~D. Corbett, A.~Kracher, , and P.~C. Canfield.
\newblock Anistropic thermodynamic and transport properties of
  single-crystalline {Ba$_{1-x}$K$_x$Fe$_2$As$_2$ ($x$ = 0 and 0.45)}.
\newblock {\em Phys.\ Rev.\ B}, 78:014507, 2008.

\bibitem{sefat_08}
A.~S. Sefat, R.~Jin, M.~A. McGuire, B.~C. Sales, D.~J. Singh, and D.~Mandrus.
\newblock {Superconductivity at 22 K in Co-Doped BaFe$_2$As$_2$ Crystals}.
\newblock {\em Phys.\ Rev.\ Lett.}, 101:117004, 2008.

\bibitem{pfisterer_80}
M.~Pfisterer and G.~Z. Nagorsen.
\newblock {\em Naturforsch.}, 35:703, 1980.

\bibitem{chu_09}
C.~W. Chu and B.~Lorenz.
\newblock High pressure studies on {Fe}-pnictide superconductors.
\newblock {\em Physica C}, 469:385, 2009.

\bibitem{canfield_09}
P.~C. Canfield, S.~L. Bud'ko, N.~Ni, J.~Q. Yan, and A.~Kracher.
\newblock {\em Phys. Rev. B}, 80:060501(R), 2009.

\bibitem{mandrus_10}
D.~Mandrus, A.~S. Sefat, M.~A. McGuire, and B.~C. Sales.
\newblock {\em Chem. Mater.}, 22:715, 2010.

\bibitem{ni_10}
N.~Ni, A.~Thaler, J.~Q. Yan, A.~Kracher, E.~Colombier, S.~L. Bud'ko, P.~C.
  Canfield, and S.~T. Hannahs.
\newblock Temperature versus doping phase diagrams for
  {Ba(Fe$_{1-x}$TM$_x$)$_2$As$_2$(TM=Ni,Cu,Cu/Co)} single crystals.
\newblock {\em Phys. Rev. B}, 82:024519, 2010.

\bibitem{kemper_09}
A.~F. Kemper, C.~Cao, P.~J. Hirschfeld, and H.-P. Cheng.
\newblock {\em Phys. Rev. B}, 80:104511, 2009.

\bibitem{bittar_11}
E.~M. Bittar, C.~Adriano, T.~M. Garitezi, P.~F.~S. Rosa,
  L.~Mendon{c}a-Ferreira, F.~Garcia, G.~de~M.~Azevedo, P.~G. Pagliuso, and
  E.~Granado.
\newblock {\em Phys. Rev. Lett.}, 107:267402, 2011.

\bibitem{wadati_10}
H.~Wadati, I.~Elfimov, and G.~A. Sawatzky.
\newblock {\em Phys. Rev. Lett.}, 105:157004, 2010.

\bibitem{rullier_albenque_09}
F.~Rullier-Albenque, D.~Colson, A.~Forget, and H.~Alloul.
\newblock {\em Phys. Rev. Lett.}, 103:057001, 2009.

\bibitem{butch_10}
N.~P. Butch, S.~R. Saha, X.~H. Zhang, K.~Kirshenbaum, R.~L. Greene, and
  J.~Paglione.
\newblock {\em Phys. Rev. B}, 81:024518, 2010.

\bibitem{fang_09}
L.~Fang, H.~Luo, P.~Cheng, Z.~Wang, Y.~Jia, G.~Mu, B.~Shen, I.~I. Mazin,
  L.~Shan, C.~Ren, and H.-H. Wen.
\newblock {\em Phys. Rev. B}, 80:140508, 2009.

\bibitem{khasanov_11}
A.~Khasanov, S.~C. Bhargava, J.~G. Stevens, J.~Jiang, J.~D. Weiss, E.~E.
  Hellstrom, and A.~Nath.
\newblock {\em J. Phys.: Condens. Matter}, 23:202201, 2011.

\bibitem{jia_95}
J.~J. Jia, T.~A. Callcott, J.~Yurkas, A.~W. Ellis, F.~J. Himpsel, M.~G. Samant,
  J.~St{\"{o}}hr, D.~L. Ederer, J.~A. Carlisle, E.~A. Hudson, L.~J. Terminello,
  D.~K. Shuh, and R.~C.~C. Perera.
\newblock First experimental results from {IBM/TENN/TULANE/LLNL/LBL} undulator
  beamline at the advanced light source.
\newblock {\em Rev.\ Sci.\ Instrum.}, 66:1394, 1995.

\bibitem{regier_07}
T.~Regier, J.~Krochak, T.~K. Sham, Y.~F. Hu, J.~Thompson, and R.~I.~R. Blyth.
\newblock Performance and capabilities of the {C}anadian {D}ragon: {T}he {SGM}
  beamline at the {C}anadian {L}ight {S}ource.
\newblock {\em Nucl. Instrum. Meth. A}, 582:93, 2007.

\bibitem{achkar_11}
A.~Achkar, T.~Regier, H.~Wadati, Y.-J. Kim, H.~Zhang, and D.~Hawthorn.
\newblock {Bulk sensitive x-ray absorption spectroscopy free of self-absorption
  effects}.
\newblock {\em Physical Review B}, 83(8):081106, 2011.

\bibitem{kurmaev_09}
E.~Z. Kurmaev, J.~A. McLeod, A.~Buling, N.~A. Skorikov, A.~Moewes, M.~Neumann,
  M~A. Korotin, Yu.~A. Izyumov, N.~Ni, and P.~C. Canfield.
\newblock Contribution of {F}e $3d$ states to the {F}ermi level of
  {C}a{F}e$_2$,{A}s$_2$.
\newblock {\em Phys. Rev. B}, 80:054508, 2009.

\bibitem{fadley_68}
C.~S. Fadley and D.~A. Shirley.
\newblock {\em Phys. Rev. Lett.}, 21(14):980, 1968.

\bibitem{blaha_01}
P.~Blaha, K.~Schwarz, G.~K.~H. Madsen, D.~Kvasnicka, and J.~Luitz.
\newblock {\em {\textbf{WIEN2k},\ An\ Augmented\ Plane\ Wave\ +\ Local\
  Orbitals\ Program\ for\ Calculating\ Crystal\ Properties}}.
\newblock Karlheinz Schwarz, Techn. Universit{\"{a}}t Wien, Austria, 2001.
\newblock {\textsc{ISBN}} 3-9501031-1-2.

\bibitem{perdew_96}
J.~P Perdew, K.~Burke, and M.~Ernzerhof.
\newblock {Generalized Gradient Approximation Made Simple}.
\newblock {\em Phys.\ Rev.\ Lett.}, 77:3865, 1996.

\bibitem{schwarz_79}
K.~Schwarz, A.~Neckel, and J.~Nordgren.
\newblock {\em J. Phys. F: Metal Phys.}, 9:2509, 1979.

\bibitem{jong_09}
S.~{de Jong}, Y.~Huang, R.~Huisman, F.~Massee, S.~Thirupathaiah, M.~Gorgoi,
  R.~Follath, J.~B. Goedkoop, and M.~S. Golden.
\newblock {High-resolution, hard x-ray photoemission investigation of
  BaFe$_2$As$_2$: Moderate influence of the surface and evidence for a low
  degree of Fe $3d$-As $4p$ hybridization of the electronic states near the
  Fermi energy}.
\newblock {\em Phys. Rev. B}, 79:115125, 2009.

\bibitem{garcia_08}
D.~R. Garcia, C.~Jozwiak, C.~G. Hwang, A.~Fedorov, S.~M. Hanrahan, S.~D.
  Wilson, C.~R. Rotundu, B.~K. Freelon, R.~J. Birgeneau, E.~Bourret-Courchesne,
  and A.~Lanzara.
\newblock {\em Phys. Rev. B}, 78:245119, 2008.

\bibitem{moulder_92}
J.~F. Moulder, W.~F. Stickle, P.~E. Sobol, and K.~Bomben.
\newblock {\em Handbook of {X}-ray photoelectron spectroscopy}.
\newblock Perkin-Elmer Corporation (Physical Electronics), 2 edition, 1992.

\bibitem{galakhov_97}
V.~R. Galakhov, A.~I. Poteryaev, E.~Z. Kurmaev, V.~I. Anisimov, St. Bartkowski,
  M.~Neumann, Z.~W. Lu, B.~M. Klein, and T.-R. Zhao.
\newblock {\em Phys. Rev. B}, 56(8):4584, 1997.

\bibitem{kurmaev_09_2}
E.~Z. Kurmaev, J.~A. McLeod, N.~A. Skorikov, L.~D. Finkelstein, A.~Moewes,
  Yu.~A. Izyumov, and S.~Clarke.
\newblock Identifying valence structure in {L}i{F}e{A}s and {N}a{F}e{A}s with
  core-level spectroscopy.
\newblock {\em J. Phys.: Condens. Matter}, 21:345701, 2009.

\bibitem{yang_09}
W.~L. Yang, A.~P. Sorini, C.-C. Chen, B.~Moritz, W.-S. Lee, F.~Vernay,
  P.~Olalde-Velasco, J.~D. Denlinger, B.~Delley, J.-H. Chu, J.~G. Analytis,
  I.~R. Fisher, Z.~A. Ren, J.~Yang, W.~Lu, Z.~X. Zhao, J.~{van den Brink},
  Z.~Hussain, Z.-X. Shen, and T.~P. Devereaux.
\newblock {\em Phys. Rev. B}, 80:014508, 2009.

\bibitem{prince_05}
K.~C. Prince, M.~Matteucci, K.~Kuepper, S.~G. Chiuzbaian, S.~Bartkowski, and
  M.~Neumann.
\newblock {\em Phys. Rev. B}, 71:085102, 2005.

\bibitem{kruger_09}
P.~Kr{\"{u}}ger.
\newblock {\em J. Phys.: Conf. Series}, 190:012006, 2009.

\bibitem{kroll_08}
T.~Kroll, S.~Bonhommeau, T.~Kachel, H.~A. D{\"{u}}rr, J.~Werner, G.~Behr,
  A.~Koitzsch, R.~H{\"{u}}bel, S.~Leger, R.~Sch{\"{u}}nfelder, A.~K. Ariffin,
  R.~Manzke, F.~M.~F. {de Groot}, J.~Fink, H.~Eschrig, B.~B{\"{u}}chner, and
  M.~Knupfer.
\newblock {\em Phys. Rev. B}, 78:220502(R), 2008.

\bibitem{bondino_09}
F.~Bondino, E.~Magnano, M.~Malvestuto, F.~Parmigiani, M.~A. McGuire, A.~S.
  Sefat, B.~C. Sales, R.~Jin, D.~Mandrus, E.~W. Plummer, D.~J. Singh, and
  N.~Mannella.
\newblock {\em Phys. Rev. Lett.}, 101:267001, 2008.

\bibitem{krause_79}
M.~O. Krause.
\newblock {Atomic Radiative and Radiationless Yields for \textit{K} and
  \textit{L} Shells}.
\newblock {\em J. Phys. Chem. Ref. Data}, 8:307, 1979.

\bibitem{kurmaev_05}
E.~Z. Kurmaev, A.~L. Ankudinov, J.~J. Rehr, L.~D. Finkelstein, P.~F. Karimov,
  and A.~Moewes.
\newblock {The L$_2$:L$_3$ intensity ratio in soft X-ray emission spectra of
  3d-metals}.
\newblock {\em J. Electr. Spectro. Relat. Phenom.}, 148:1, 2005.

\bibitem{kurmaev_08}
E.~Z. Kurmaev, R.~G. Wilks, A.~Moewes, N.~A. Skorikov, Yu.~A. Izyumov, L.~D.
  Finkelstein, R.~H. Li, and X.~H. Chen.
\newblock X-ray spectra and electronic structures of the iron arsenide
  superconductors {\textit{r}}{FeAsO}$_{1-x}${F}$_x$ ({\textit{r}} = {La,Sm}).
\newblock {\em Phys. Rev. B}, 78:220503(R), 2008.

\bibitem{anisimov_09}
V.~I. Anisimov, E.~Z. Kurmaev, A.~Moewes, and I.~A. Izyumov.
\newblock Strength of correlations in pnictides and its assessment by
  theoretical calculations and spectroscopy experiments.
\newblock {\em Physica C}, 469:442, 2009.

\bibitem{grioni_92}
M.~Grioni, J.~F. {van Acker}, M.~T. Czy{\u{z}yk}, and J.~C. Fuggle.
\newblock {\em Phys. Rev. B}, 45:3309, 1992.

\bibitem{shein_08}
I.~R. Shein and A.~L. Ivanovskii.
\newblock {Electronic Structure of New Oxygen-Free 38-K Superconductor
  Ba$_{1-x}$K$_x$Fe$_2$As$_2$ in Comparison with BaFe$_2$As$_2$ from the First
  Principles}.
\newblock {\em JETP\ Letters}, 88:107, 2008.

\bibitem{freelon_10}
B.~Freelon, Y.-S. Liu, C.~R. Rotundu, S.~D. Wilson, J.~Guo, J.-L. Chen,
  W.~Yang, C.~Chang, P.~A. Glans, P.~Shirage, A.~Iyo, and R.~J. Birgeneau.
\newblock {\em Journal of the Physical Society of Japan}, 79(7):074716, 2010.

\bibitem{singh_08}
D.~J. Singh.
\newblock Electronic structure and doping in {BaFe$_2$As$_2$} and {LiFeAs}:
  {D}ensity functional calculations.
\newblock {\em Phys.\ Rev.\ B}, 78:094511, 2008.

\bibitem{sekiba_09}
Y.~Sekiba, T.~Sato, K.~Nakayama, K.~Terashima, P.~Richard, J.~H. Bowen,
  H.~Ding, Y.-M. Xu, L.~J. Li, G.~H. Cao, Z.-A. Xu, and T.~Takahashi.
\newblock {\em New J. Phys.}, 11:025020, 2009.

\bibitem{olariu_11}
A.~Olariu, F.~Rullier-Albenque, D.~Colson, and A.~Forget.
\newblock {\em Phys. Rev. B}, 83:054518, 2011.

\end{thebibliography}

\end{document}